\documentclass[journal,twoside,web]{ieeecolor}
\usepackage{tmi}
\usepackage{cite}
\usepackage{amsmath,amssymb,amsfonts}
\usepackage{algorithmic}
\usepackage{graphicx}
\usepackage{textcomp}
\usepackage{orcidlink} 
\usepackage{times}
\usepackage{epsfig}
\usepackage{graphicx}
\usepackage{amsmath}
\usepackage{amssymb}
\usepackage[ruled,vlined,linesnumbered]{algorithm2e}
\usepackage{booktabs}
\usepackage{caption}
\usepackage{subcaption}
\usepackage{pgfplots}
\pgfplotsset{compat=1.16}
\usepackage{comment}
\usepackage{array, makecell} 
\usepackage{xcolor}
\usepackage{lettrine}
\usepackage{physics} 
\usepackage{acronym}
\acrodef{LBP}[LBP]{Low Back Pain}
\acrodef{2D}[2D]{Two-Dimensional}
\acrodef{FCN}[FCN]{Fully Convolutional Network}
\acrodef{GAME}[GAME]{Grid Average Mean Absolute Error}
\acrodef{DL}[DL]{Deep Learning}
\acrodef{US}[US]{Ultrasound}
\acrodef{CT}[CT]{Computerized Tomography}
\acrodef{TrA}[TrA]{Transversus Abdominis}
\acrodef{ICC}[ICC]{Intraclass Correlation Coefficient}
\acrodef{CNN}[CNN]{Convolutional Neural Network}
\acrodef{JCU}[JCU]{James Cook University}
\acrodef{ADIM}[ADIM]{Abdominal Drawing-In Maneuver}
\acrodef{MWA}[MWA]{Muscle Width Agreement}
\acrodef{MAE}[MAE]{Mean Absolute Error}
\acrodef{LCFCN}[LCFCN]{Localization-based Counting Fully Convolutional Network}
\usepackage[switch]{lineno}
\usepackage{everypage}
\usepackage{tikz}
\usepackage{lipsum}

\newcommand{\alz}[1]{\textcolor{black}{#1}}

\usepackage{moreverb,url}

\usepackage[colorlinks,bookmarksopen,bookmarksnumbered,citecolor=red,urlcolor=red]{hyperref}

\newcommand\BibTeX{{\rmfamily B\kern-.05em \textsc{i\kern-.025em b}\kern-.08em
T\kern-.1667em\lower.7ex\hbox{E}\kern-.125emX}}

\AddEverypageHook{%
 \begin{tikzpicture}[remember picture,overlay, scale=0.6]
 \node[
    scale=0.63,
    opacity=1,
    xshift=0,
    yshift=0,
    color=black
    ]
    at (14, 3.5)  
    {\parbox{24cm}{%
    \begin{tabular}{c}
      This is the author's version of an article that has been published in this journal. Changes were made to this version by the publisher prior to publication. \\
      The final version of record is available at
         \textcolor{blue}{https://doi.org/10.1109/JBHI.2021.3085019}
    \end{tabular}}};
\end{tikzpicture}
}

\AddEverypageHook{%
 \begin{tikzpicture}[remember picture,overlay, scale=0.6]
 \node[
    scale=0.63,
    opacity=1,
    xshift=0,
    yshift=0,
    color=black
    ]
    at (13.5, -41.9)  
    {Copyright (c) 2021 IEEE. Personal use is permitted. For any other purposes, permission must be obtained from the IEEE by emailing pubs-permissions@ieee.org.};
\end{tikzpicture}
}

\def\BibTeX{{\rm B\kern-.05em{\sc i\kern-.025em b}\kern-.08em
    T\kern-.1667em\lower.7ex\hbox{E}\kern-.125emX}}
\begin{document}


\title {A Deep Learning Localization Method for Measuring Abdominal Muscle Dimensions in Ultrasound Images}
\author{Alzayat Saleh\textsuperscript{\orcidlink{0000-0001-6973-019X}}, Issam H. Laradji\textsuperscript{\orcidlink{0000-0002-9713-3269}}, Corey Lammie\textsuperscript{\orcidlink{0000-0001-5564-1356}}, David Vazquez\textsuperscript{\orcidlink{0000-0002-2845-8158}}, Carol A Flavell\textsuperscript{\orcidlink{0000-0003-0135-3425}}, and Mostafa~Rahimi~Azghadi\textsuperscript{\orcidlink{0000-0001-7975-3985}}
\thanks{Manuscript received January 26, 2021. \\
Study approval was obtained from the Human Research Ethics Committee of the James Cook University (JCUH7878).\\
The author(s) received no financial support for the research,
authorship, and/or publication of this article.\\
The author(s) declared no potential conflicts of interest with
respect to the research, authorship, and/or publication of this
article.\\
(Corresponding author: Alzayat Saleh and Mostafa Rahimi Azghadi.)}
\thanks{A. Saleh is with College of Science and Engineering, James Cook University, Townsville, Australia (e-mail:\href{mailto:alzayat.saleh@my.jcu.edu.au}{alzayat.saleh@my.jcu.edu.au}).}
\thanks{ I. H. Laradji is with McGill University, Montreal, Canada. and Element AI, Montreal, Canada (e-mail: {\href{mailto:issam.laradji@gmail.com}{issam.laradji@gmail.com}}).}
\thanks{C. Lammie is with College of Science and Engineering, James Cook University, Townsville, Australia (e-mail: {\href{mailto:corey.lammie@jcu.edu.au}{corey.lammie@jcu.edu.au}}).}
\thanks{D. Vazquez is with Element AI, Montreal, Canada (e-mail: {\href{mailto:david.vazquez@servicenow.com}{david.vazquez@servicenow.com}}).}
\thanks{C. A. Flavell is with College of Healthcare Sciences, James Cook University, Townsville, Australia (e-mail: {\href{mailto:carol.flavell@jcu.edu.au}{carol.flavell@jcu.edu.au}}).}
\thanks{M. Rahimi Azghadi is with College of Science and Engineering, James Cook University, Townsville, Australia (e-mail: {\href{mailto:mostafa.rahimiazghadi@jcu.edu.au}{mostafa.rahimiazghadi@jcu.edu.au}}).}}

\maketitle

\begin{abstract}
Health professionals extensively use \ac{2D} \ac{US} videos and images to visualize and measure internal organs for various purposes including evaluation of muscle architectural changes. \ac{US} images can be used to measure abdominal muscles dimensions for the diagnosis and creation of customized treatment plans for patients with \ac{LBP}, however, they are difficult to interpret. Due to high variability, skilled professionals with specialized training are required to take measurements to avoid low intra-observer reliability. This variability stems from the challenging nature of accurately finding the correct spatial location of measurement endpoints in abdominal \ac{US} images. In this paper, we use a \ac{DL} approach to automate the measurement of the abdominal muscle thickness in \ac{2D} \ac{US} images. By treating the problem as a localization task, we develop a modified \ac{FCN} architecture to generate blobs of coordinate locations of measurement endpoints, similar to what a human operator does. We demonstrate that using the TrA400 \ac{US} image dataset, our network achieves a \ac{MAE} of 0.3125 on the test set, which almost matches the performance of skilled ultrasound technicians. Our approach can facilitate next steps for automating the process of measurements in \ac{2D} \ac{US} images, while reducing inter-observer as well as intra-observer variability for more effective clinical outcomes.
\end{abstract}


\raggedbottom 

\section{Introduction}
\IEEEPARstart{H}{ealth} professionals extensively use \ac{2D} Ultrasound (US) videos and images to visualize and measure internal organs for various purposes including evaluation of muscle architectural and functional changes for diagnostic purposes, and as an outcome measure to evaluate rehabilitation effects. 
Compared to \ac{CT} scans, \ac{US} imaging is low-cost \cite{so2011medical}, more sensitive \cite{van2011comparison}, and does not expose patients to ionizing radiation \cite{vera1997assessment}.

\begin{figure*}[!t]
\centering
\includegraphics[width=0.7\textwidth]{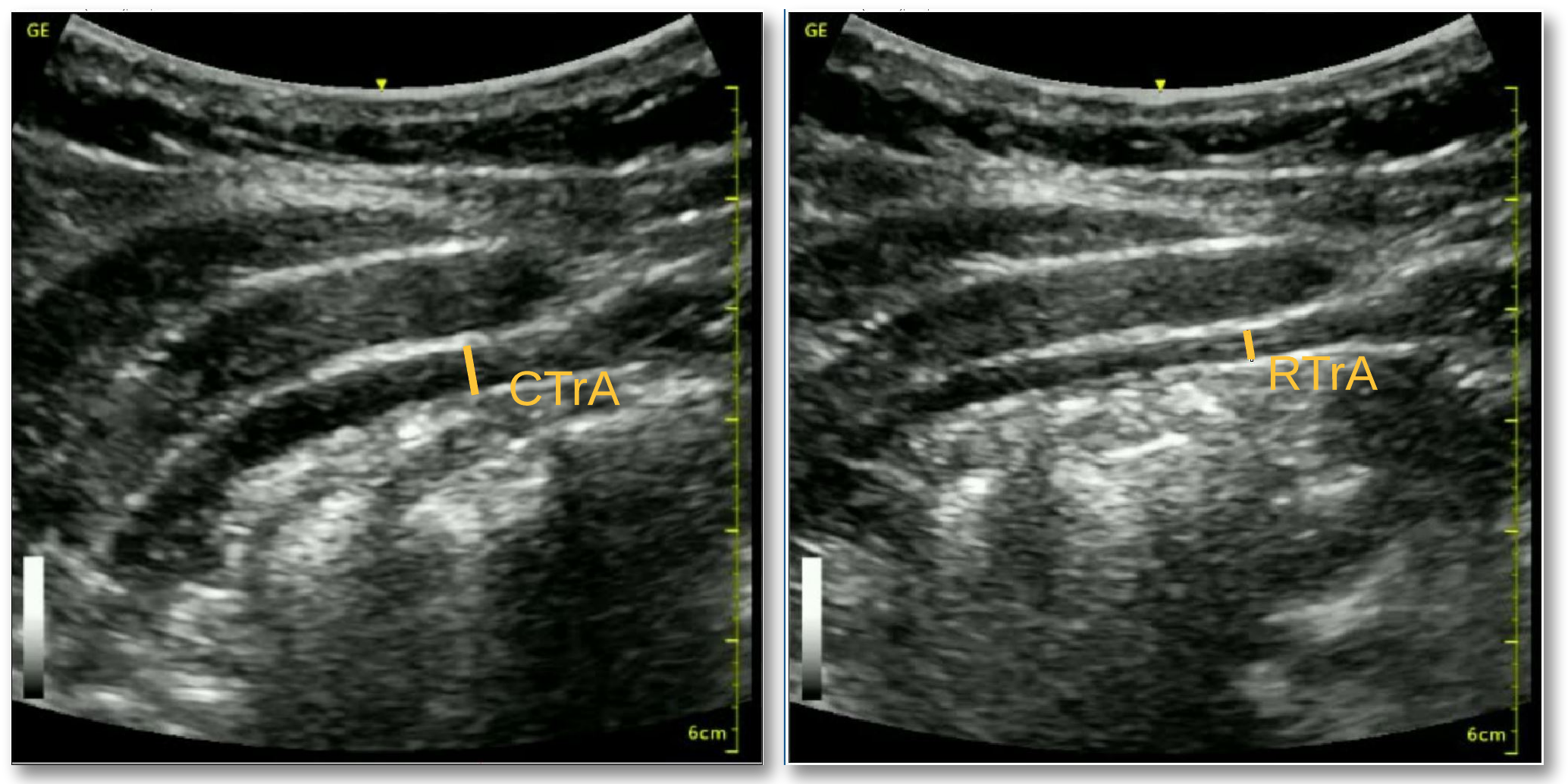}
\caption{Images depicting \ac{TrA} measurements, in contracted (left) and resting (right) states.}
\label{fig:sample}
\end{figure*}

Two-dimensional \ac{US} imaging is a valid method of measuring abdominal muscle dimensions in patients undergoing physical rehabilitation \cite{koppenhaver2009rehabilitative}.
Previous literature~\cite{kennedy2019intra} has demonstrated that the measurement of the \ac{TrA}, a broad paired muscular sheet found on the lateral sides of the abdominal wall, can be used to quantitatively express dysfunctional \ac{TrA} activation. In terms of calculating \ac{TrA} activation (TrA-C), US measurements of \ac{TrA} thickness at rest (RTrA), and when fully contracted (CTrA) have been used as a preeminent measurement~\cite{kennedy2019intra, flavell2019measurement}. Some examples of these measurements are shown in Fig.~\ref{fig:sample}. 

To evaluate the intra-rater and inter-rater reliability of measurements of \ac{TrA} muscle thickness conducted by novice examiners, the authors of \cite{hoppes2015ultrasound} conducted a study on military personnel examined by two ultrasound technicians and reported intra-rater \ac{ICC} values from 0.90 to 0.98. The inter-rater \ac{ICC} values ranged from 0.39 to 0.79 and concluded that the inter-rater reliability of two novice examiners was poor to fair. This low reliability between examiners challenges the efficacy of this method of muscle measurement in diagnostics and rehabilitation. 

Reliability of US-based \ac{TrA} measurement is deemed to be dependent on  standardized methodological factors such as imaging technique, or issues associated with landmark detection on the \ac{2D} raw image \cite{koppenhaver2009rehabilitative}. An automated measurement tool that decreases variability at the image measurement stage may help health professionals make more accurate diagnoses and evaluations of rehabilitation progress. In addition, image measurement may be expedited by the tool, providing benefit in terms of saving clinical time and cost and improving patient outcomes. \alz{\acf{DL}} offers a great potential in developing such tool, and is therefore investigated in this paper.

Previous \ac{DL} works on \ac{US}~\cite{sofka2017fully,gilbert2019automated} have framed the measurement task as a landmark detection problem, where the goal is to identify the key points on the image. In the general landmark detection approach (e.g. facial landmark detection), there is a defined local appearance of these landmarks. However, the measurements of \ac{TrA} muscle thickness from \ac{US} images is defined from local appearance and global structural information of the muscle. Consequently, we frame our task as localization of the caliper endpoints, similar to what a \ac{US} technician does, to increase user-interpretability and use of global structural information. 

For target localization in images and for many other computer vision tasks, \ac{CNN}\cite{o2015introduction,albawi2017understanding,yamashita2018convolutional} has become the dominant \ac{DL} solution. The standard \ac{CNN} is not fully convolutional because it often contains dense layers\cite{lin2013network}, which have many parameters (i.e. computationally expensive). 
In contrast, \ac{FCN}\cite{long2015fully} is an architecture employing only convolution layers. Avoiding the use of dense layers means fewer parameters (i.e. faster to train the network). 
\ac{FCN} consists of a downsampling path, used to extract and interpret the context, and an upsampling path, which allows for localization of each pixel. 
\ac{FCN} also uses skip connections\cite{drozdzal2016importance} to retrieve the fine-grained spatial information missed in the downsampling path.

\begin{figure*}[!t]
\centering
\includegraphics[width=\textwidth]{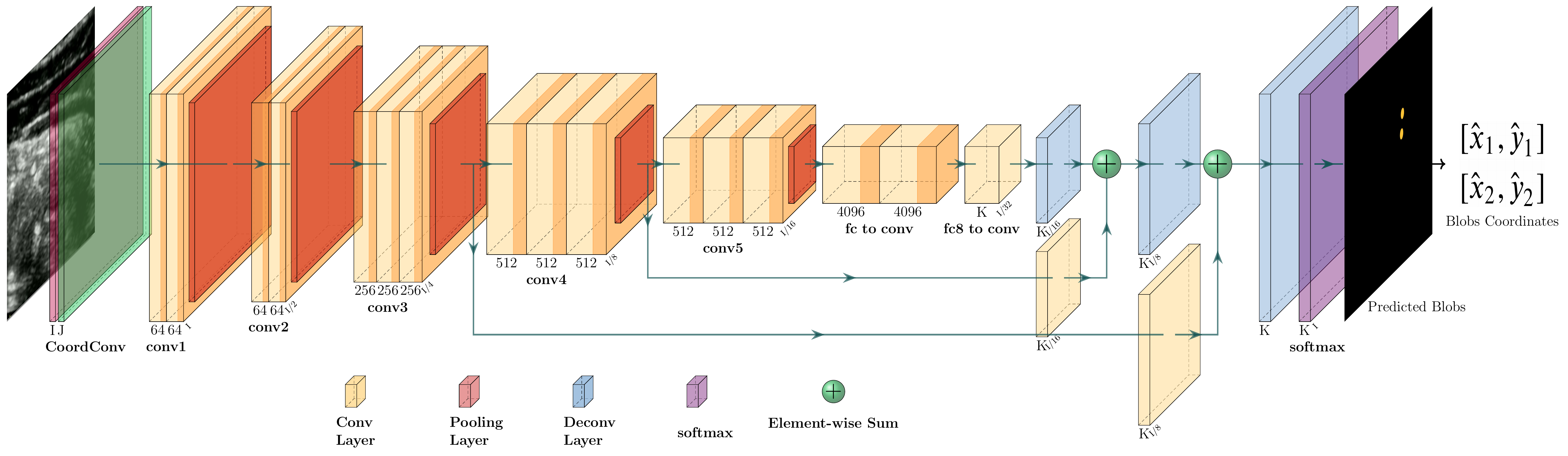}
\caption{\textbf{CoordConv-FCN8 architecture.} 
The first layer is CoordConv followed by the VGG16 backbone, which is used to extract features from the input image. 
The second component is the decoder that outputs per-pixel scores. 
Here, the two highlighted blobs are predicted locations for the targeted measurement.}
\label{fig:model}
\end{figure*}

\alz{
In this paper, an object localization framework resembling the ultrasound technicians procedure is proposed to solve the problem of measurement of abdominal muscle thickness in \ac{2D} \ac{US} images. Our specific contributions are as follows:
\begin{enumerate}
\item We propose a novel localization framework that uses FCN8~\cite{long2015fully} with  a CoordConv layer~\cite{liu2018intriguing}  to detect the points for muscle dimension measurement while leveraging \ac{LCFCN} loss to maximize the performance of the network with point-level supervision;
\item We demonstrate that our novel approach of using \ac{LCFCN} along with CoordConv yields significant improvement over conventional landmark detection and almost matches the performance of ultrasound technicians.
\end{enumerate}
}

\section{Related Work} \label{sec:relwork}
\ac{DL} has been extensively used in various medical applications including \ac{US} image processing. For instance, it has been used for automated identification and segmentation of tumours in breast ultrasound images~\cite{xian2018automatic,shin2018joint,chiang2018tumor}. In addition, it has been employed for 
prenatal screening, which is a classic therapeutic application of \ac{US}~\cite{timor1989use,cameron2009prenatal}. The diagnosis of possible fetal problems or chromosomal abnormalities is done through monitoring fetal growth and development. This diagnosis can be complicated, requiring a highly trained doctor to promptly recognise the scan planes and structures of interest.
Baumgartner \emph{et al.}~\cite{baumgartner2017sononet} used \ac{DL} to significantly simplify this process by automated fetal standard scan plane for real-time localization and detection in these planes in freehand ultrasound. 
\alz{Cronin  \emph{et al.}~\cite{cronin2020fully}  trained a deep neural networks (based on U-net) to detect muscle fascicles and aponeuroses using a set of labelled musculoskeletal ultrasound images. For ultrasound muscle segmentation, Cunningham \emph{et al.}~\cite{cunningham2016real} have developed a method for segmenting five bilateral cervical muscles and the spine via ultrasound images.}
\ac{DL} methods have also proven to outperform traditional methods in many other ultrasound imaging applications~\cite{milletari2016v,ronneberger2015u}.

Conventional landmark detection has so far been the dominant method in US localization approaches~\cite{sofka2017fully,gilbert2019automated,payer2016regressing,jg2018deep,liu2020deep}. For instance, fusing local appearance of each landmark in an ultrasound image with the spatial configuration of the rest of the landmarks has achieved good landmark localization in a previous work~\cite{payer2016regressing}. 
The standard landmark detection approach has been used in~\cite{sofka2017fully} to detect measurement keypoint locations for left ventricular internal dimension measurements by computing their regression estimates using a centre of mass layer with a \ac{CNN}. Gilbert  \emph{et al.}\cite{gilbert2019automated} extended the work of Sofka  \emph{et al.}\cite{sofka2017fully} by targeting the intra-ventricular septum and the left ventricular posterior wall measurements. They introduced anatomically meaningful ground truth heatmaps, which follow the expected spatial distribution of keypoints.

The localization approach has been also used by Sloun  \emph{et al.}~\cite{jg2018deep} to develop a \ac{CNN} that automatically identifies clinically relevant B-line artifacts in lung from an ultrasound video, offering concurrent localization by calculating neural attention maps. Similarly, Liu  \emph{et al.}\cite{liu2020deep} proposed a method implemented based on a modified sub-pixel \ac{CNN} architecture, termed as SPCN-ULM.
However, to the best of our knowledge, no previous works have utilized an \ac{FCN} variants such as an \alz{LCFCN or LCFCN+CoordConv} for US image muscle thickness measurements.

\section{Methods}
In this section we describe the methods used to develop and train our network architecture depicted in Fig.~\ref{fig:model}.

\subsection{Network Architecture}
Our novel network architecture is largely based on the FCN8~\cite{long2015fully} network architecture, which is one of the most popular \ac{DL} semantic segmentation network architectures demonstrating significant performance in a variety of segmentation tasks~\cite{fayyaz2016stfcn,nayem2020lulc,zhao2018semantic}. The complete architecture consists of a segmentation network based on FCN8 with an ImageNet~\cite{ILSVRC15} pre-trained VGG16~\cite{Simonyan2014VGG} backbone. 
\alz{To overcome the problem of VGG-16 overfitting to the small available dataset, we used augmentation and regularization. Also, cross-validation was used to report performance across 10 folds to confirm the consistency of the results.}
All layers are sequenced with the ReLU activation function, and batch normalization is included between convolutional blocks for regularization. Standard Dropout layers are not used, as FCN8 is fully convolutional and neighbouring pixels in images wihin the TrA400 dataset are strongly correlated. Instead, SpatialDropout~\cite{tompson2015efficient} layers are used in-place, which restrict the co-adaptation of pixels with their neighbouring pixels across feature maps. This approach has previously been empirically verified by Lee  \emph{et al.}~\cite{lee2020revisiting}, which reported a significant performance improvement when replacing standard Dropout layers with SpatialDropout layers.

\subsection{CoordConv Layer}
\begin{figure*}[!t]
\includegraphics[width=0.8\textwidth]{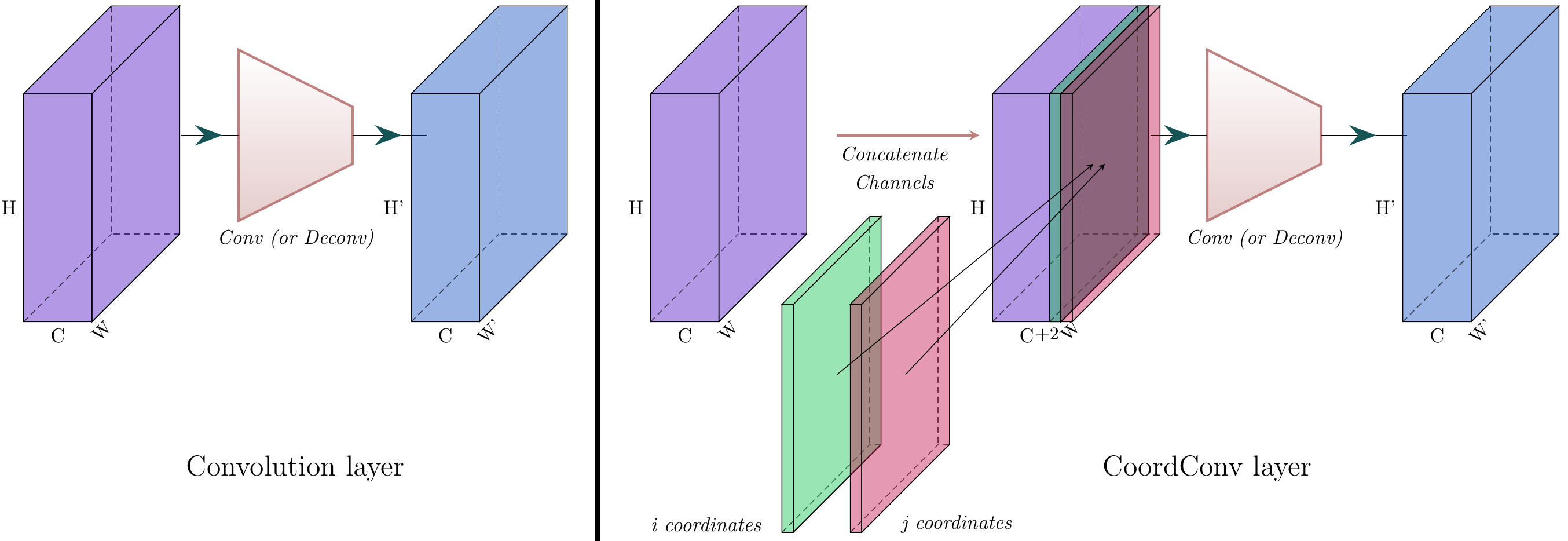}
\centering
\caption{\alz{A visual representation of how CoordConv differs from the standard convolution layer.
The left is a schematic diagram of standard convolutional layer maps from a representation block with shape $h \times w \times c $ to a new representation of shape $h' \times w' \times c'$ . 
The right schematic diagram of CoordConv layer shows that it differs from the standard convolutional layer by first concatenating extra hard-coded $i$ and $j$ coordinates channels to the incoming representation.}}
\label{fig:cord}
\end{figure*}

\alz{To locate caliper endpoints using pixel-wise spatial location information, a CoordConv~\cite{liu2018intriguing} layer was added to the beginning of the FCN8~\cite{long2015fully} network architecture for a more effective transformation of the positional information. The CoordConv layer is a simple extension to the standard convolutional layer, which adds two channels to the input. The first channel is row-wise with $i$ coordinates ($h \times w$ rank-1 matrix), while the second channel is column-wise with $j$ coordinates. Fig.~\ref{fig:cord} provides a visual representation of CoordConv. Here, $h$ and $w$ are the input image $\mathbf{X}$ height and width, respectively, given $\mathbf{X} \in \mathbb{R}^{h \times w \times c}$, where $c$ is the number of channels.
These two extra channels allow the \ac{CNN} to learn the order of convolutional filters to know where they are in Cartesian space through coordinates to properly learn translation equivariance~\cite{liu2018intriguing}.
Translational Equivariance\cite{celledoni2021equivariant} is a very crucial attribute of the \ac{CNN} where the object location in the image does not need to be fixed in order to be detected by the \ac{CNN}.
This allows learned feature detectors to detect features regardless of their position (though not their orientation or scale) in an image. 
The CoordConv layer keeps the degree to which Translational Equivariance is learned. If weights from the coordinates learn to become zero, CoordConv behaves like standard convolution. On the other hand, if translation dependence is useful for the downstream task, as it is the case in our task, it will be properly learned. Moreover, the positional information carried by CoordConv layers acts as “soft constraint” and guarantees that the segmented blobs are around the correct location of caliper points.
}

\subsection{LCFCN Loss}
\label{sec:lcfcn}
\alz{The LCFCN loss function consists of four different sub-losses. The first two sub-losses, the image-level and the point-level loss, drive the model to predict the semantic segmentation labels for each pixel in the image. The other two sub-losses, i.e. split-level and false-positive, encourage the model to output a single blob for each object instance and eliminate blobs that have no object instances. The LCFCN loss function only requires point-level annotations that define the locations of the objects rather than their sizes, and shapes.}

\begin{figure*}[!t]
\includegraphics[width=0.99\textwidth]{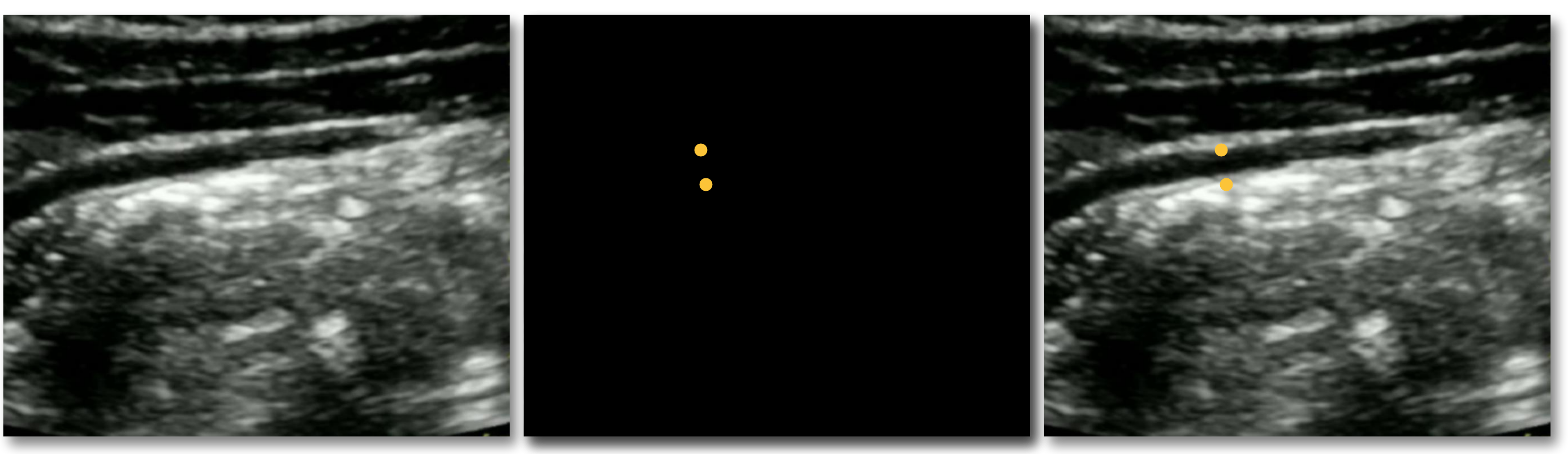}
\centering
\caption{Point annotations in a sample \ac{US} image ($\mathbf{X}$) (left). The points in the training mask image ($\mathbf{Y}$) (middle) are extracted from a human-annotated image (right) and are inflated and highlighted
for visibility, but only the center pixel and its class label are
collected and used.}
\label{fig:annotate}
\end{figure*}

This loss function was used to learn a single blob per caliper endpoint in the image using point-level annotations (Fig.~\ref{fig:annotate}). While LCFCN was originally designed for counting, it is also able to locate objects and segment them~\cite{laradji2019instance,laradji2020weaklyWS,laradji2020looc,laradji2019masks,saleh2020realistic}, by refining the activation output that determines the likelihood of a pixel belonging to the localization or segmentation target. In our localization task, we obtain per-pixel probabilities by applying the Softmax activation function to compute $\hat{Y}$, which contains the likelihood that a pixel either belongs to the background or muscle edge. The LCFCN loss $\mathcal{L}_L$ is described using equation (\ref{eq:loss_lcfcn})
\begin{equation}
\begin{split}
\mathcal{L}_L  &= \underbrace{\mathcal{L}_I(\hat{Y},Y)}_{\text{Image-level loss}} + \underbrace{\mathcal{L}_P(\hat{Y},Y)}_{\text{Point-level loss}}\\ &+ \underbrace{\mathcal{L}_S(\hat{Y},Y)}_{\text{Split-level loss}} + \underbrace{\mathcal{L}_F(\hat{Y},Y)}_{\text{False positive loss}}\;,
\label{eq:loss_lcfcn}
\end{split}
\end{equation}
where $Y$ represents the point annotation ground-truth. 

\alz{The \textbf{image-level loss} ($\mathcal{L}_I$) trains the model to predict whether there is an object in the image, and it is described using equation (\ref{eq:imagelevel}).
\begin{equation}
\mathcal{L}_I(\hat{Y}, Y) = -\frac{1}{|C_e|}\sum_{c\in C_e}\log(\hat{Y}_{t_cc}) -\frac{1}{|C_{\neg{e}}|}\sum_{c\in C_{\neg{e}}}\log(1 - \hat{Y}_{t_cc}) \;,
\label{eq:imagelevel}
\end{equation}
where $t_c = \text{argmax}_{i \in \mathcal{I}} \hat{Y}_{ic}$. Here, each class $c \in C_e$ which shows the set of classes present in the image, while $C_{\neg{e}}$ is the set of classes not present in the image.}

\vspace{0.1cm}
\alz{The \textbf{point-level loss} ($\mathcal{L}_P$) encourages the model to predict a pixel for each object instance, and it is described using equation (\ref{eq:pointlevel}).
\begin{equation} \label{eq:pointlevel}
\mathcal{L}_P(\hat{Y}, Y) = -\sum_{i\in \mathcal{I}_s}\log(\hat{Y}_{iY_i})\;,
\end{equation}
where $Y_i$ represents the true label of pixel $i$ and $\mathcal{I}_s$ represents the locations of the object instances.}

\vspace{0.1cm}
\alz{The \textbf{split-level loss }($\mathcal{L}_S$) enforces the model to predict a single blob per instance~\cite{laradji2018blobs}, and it is described using equation (\ref{eq:edgelevel}).
\begin{equation}
\begin{split}
\mathcal{L}_S(\hat{Y}, Y) &=  -  \sum_{i \in T_b} \alpha_i \log(\hat{Y}_{i0}),
\end{split}
\label{eq:edgelevel}
\end{equation}
where $\hat{Y}_{i0}$ is the probability that pixel $i$ belongs to the background class, $\alpha_i$ is the number of point-annotations in the blob in which pixel $i$ lies, and $T_b$ is the set of pixels representing the boundaries determined by the local and global segmentation.}

\vspace{0.1cm}
\alz{The \textbf{false-positive loss} ($\mathcal{L}_F$) discourages the model from predicting a blob with no point annotations, in order to reduce the number of false positive predictions, and is described using equation (\ref{eq:falsepositive}).
\begin{equation}
\mathcal{L}_F(\hat{Y}, Y) = - \sum_{i \in B_{fp}} \log(\hat{Y}_{i0}),
\label{eq:falsepositive}
\end{equation}
where $B_{fp}$ is the set of pixels constituting the blobs predicted for each class (except the background class) that contain no ground-truth point annotations (note that $S_{i0}$ is the probability that pixel $i$ belongs to the background class).}

Applying LCFCN loss on the original activation map generates a small blob around the centre of the object, in our case the caliper endpoints. To measure the \ac{TrA} thickness, we take the centroid of each blob to get coordinate locations and measure the distances between the coordinate pairs that our model predicted.

\subsection{TrA400 Dataset}
Ethics approval for the use of the ultrasound image set in this study was granted by the Human Research Ethics Committee of \ac{JCU} (JCUH7878).
The TrA400 dataset consists of 400 \ac{2D} \ac{US} images used to measure the thickness and activation changes in the \ac{TrA}. The \ac{US} images were captured from abdominal
\ac{US} videos using methods described previously in Kennedy  \emph{et al.}~\cite{kennedy2019intra}.
\alz{
The 400 \ac{2D} \ac{US} images are from 56 participants and at least 4 images per participant were captured as detailed below. For the first 38 participants, each one was examined by two examiners each taking 2 types of measurements, “freehand’’ and ‘‘probe force device’’ for two different states, i.e. CTrA and RTrA. Therefore, $38\times2\times2\times2=304$ images were captured. For some of these 38 participants, 20 additional random measurements were also performed, bringing the total number of measurements to 324. For the remaining 18 participants, 4 measurements per participant were captured in CTrA and RTrA and using “freehand’’ and ‘‘probe force device’’. For 10 of these 18 participants 2 extra measurements were also randomly conducted. These resulted in 82 images from the 18-subject cohort, bringing the total number of images in our collected dataset to 406 images. Two of these images were unreadable by humans so were removed. In addition, an additional 4 images were removed randomly for the built dataset to reach 400 still images to carry out cross-validation across 10 folds.
}

\ac{US} videos were imported into VideoPad\textsuperscript{\textregistered} (NCH Software, 2016). The examiner extracted one still image from each imported video. Still resting thickness of the TrA (RTrA) images from all videos were captured at the end-expiratory phase and contracted thickness of the TrA (CTrA) images were taken during \ac{ADIM}~\cite{kennedy2019intra}. 
\alz{The end-expiratory phase and contracted images were extracted from different videos.} The examiner selected still images based on the highest quality visualization of TrA available. 

\alz{Several videos were captured per participant, during the US session. Participants remained on the treatment plinth for the duration of video capture. This gave the examiner a good opportunity to select a proper still image from each of the videos.}

Full details regarding the study design and inclusion and exclusion criteria of participants are available in previous studies~\cite{kennedy2019intra,flavell2019measurement}.

\begin{figure}[!t]
\includegraphics[width=0.48\textwidth]{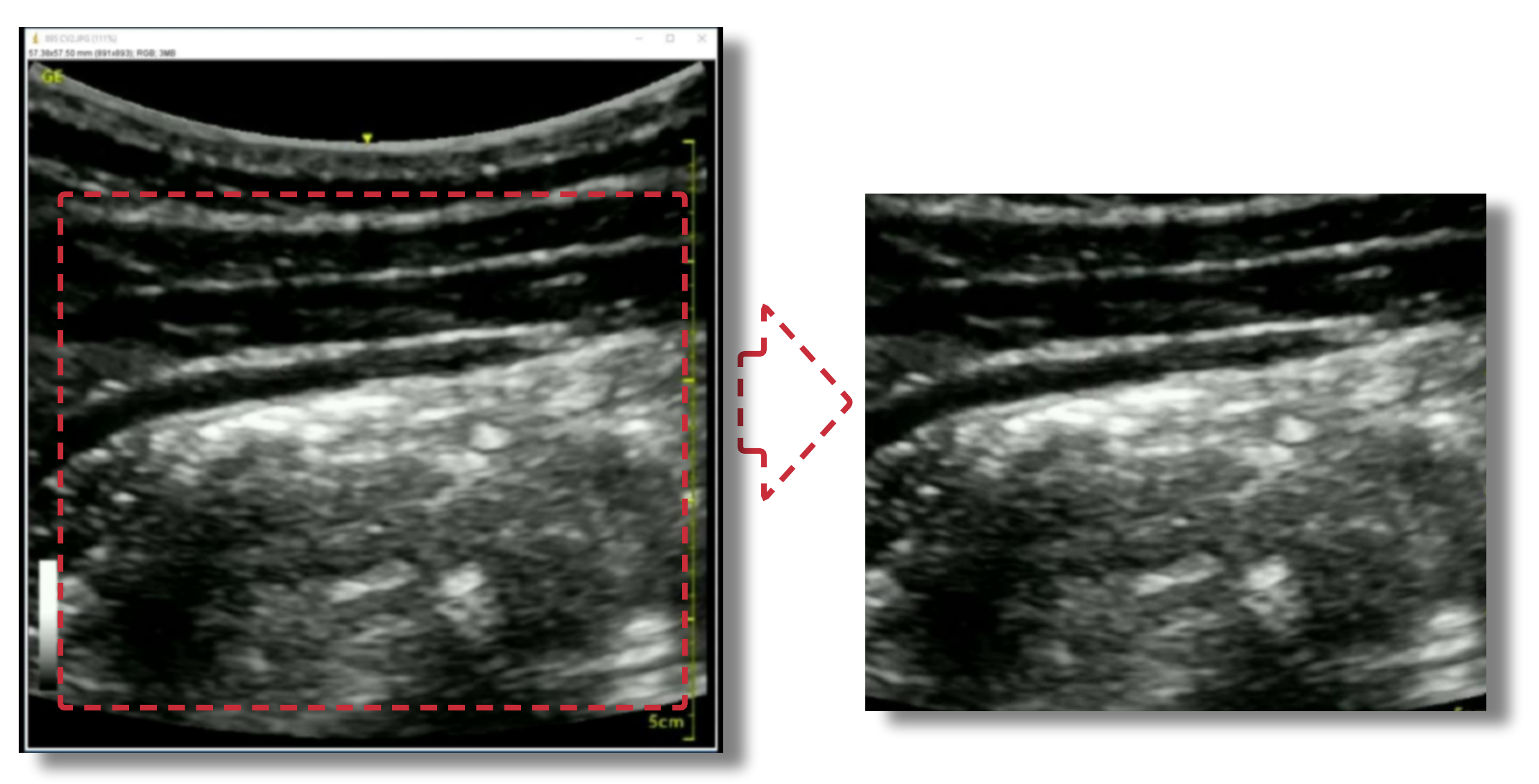}
\centering
\caption{A sample full resolution \ac{US} image (left) and its corresponding (right) cropped image ($895 \times 745$ pixels).}
\label{fig:crop}
\end{figure}

\section{Experiments}
In this section we describe the experiments performed, and provide details of the methods and evaluation metrics used in our experiments.

\begin{figure}[!b]
\includegraphics[width=0.48\textwidth]{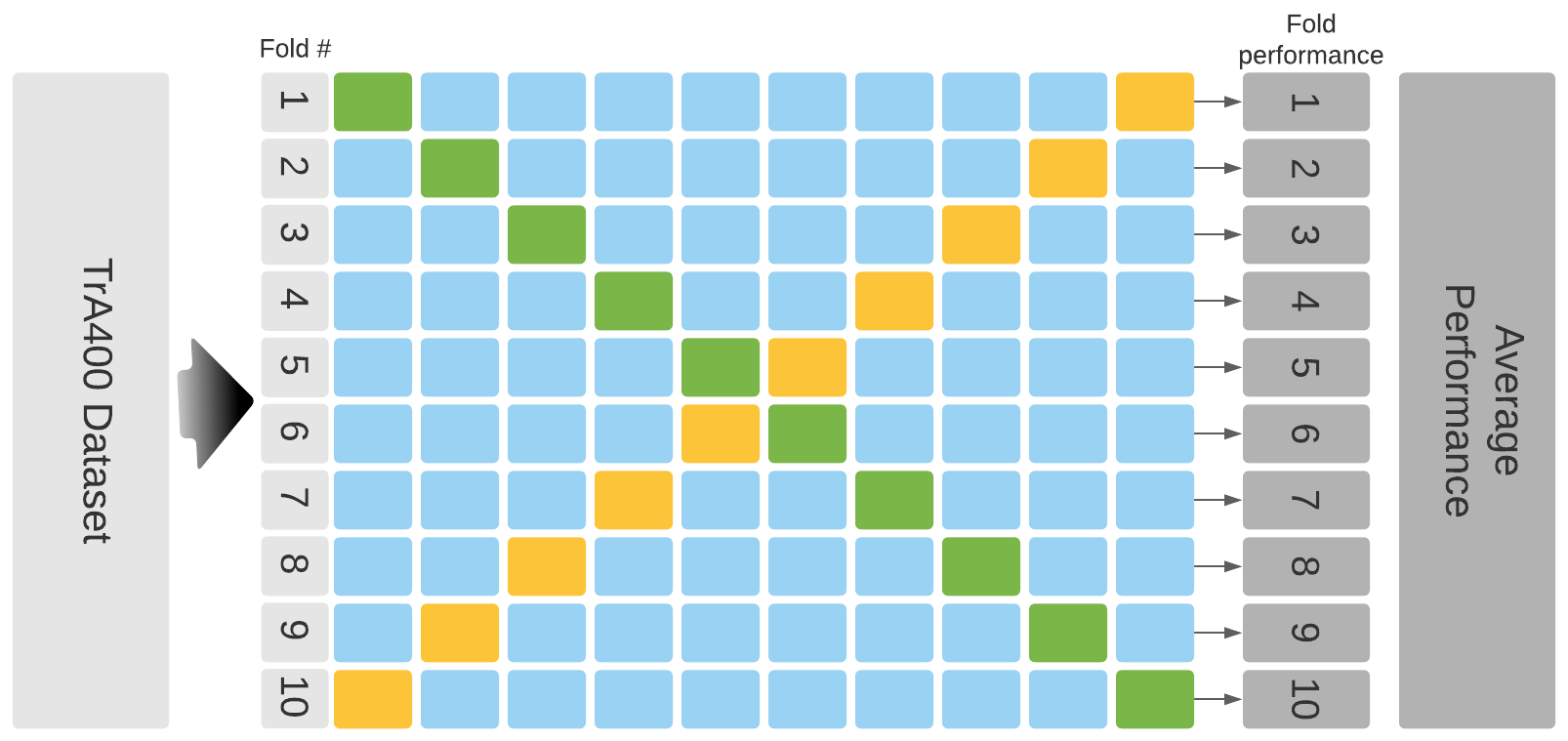}
\centering
\caption{Cross-validation distribution for TrA400 dataset. Each rectangle represents a group of 40 images. Green rectangles are used to denote test splits, yellow rectangles are used to show validation splits, and blue rectangles are used to demonstrate train splits. Performance is reported across all folds.}
\label{fig:crossval}
\end{figure}

\subsection{Dataset Preparation}
The extracted US images were annotated by an examiner to measure the muscle thickness. 
\alz{The caliper endpoints were defined for abdominal muscle dimension measurements as the start and the endpoints of the line in the middle of the TrA, which is drawn by a human examiner between the superior aspect of the inferior hyperechoic fascial border of the TrA and the inferior border of its superior hyperechoic fascial line, to measure the muscle thickness, as depicted in Fig.~\ref{fig:sample}.}
In order to train our deep learning network, point-level annotated binary masks were obtained by converting the start and the end of the line in the middle of the TrA to ($x_1 y_1$) and ($x_2, y_2$) coordinates, respectively. Here, ($x_i y_i$) coordinates have been given a value one as foreground, while the rest of the image pixels are turned to zeros as the background. The whole TrA400 dataset consists of precise image-mask pairs ($\mathbf{X}$, $\mathbf{Y}$) (see Fig.~\ref{fig:annotate}). The extracted US images original size are between  $1010 \times 1040$ pixels to $898 \times 945$ pixels. However, the region of interest  for the \ac{TrA} measurements is within a $895 \times 745$ pixels range. Therefore, we cropped the images to $895 \times 745$ pixels region as shown in Fig.~\ref{fig:crop}. \alz{We manually defined the region of interest to exclude the unwanted parts of the images that included markers and text, which were positioned outside the region of interest, i.e. the \ac{TrA} muscle in the middle of the image.}

All 400 images of the TrA400 dataset were split into 10 folds of 3 subsets including training set ($n=320$), validation set ($n=40$), and test set ($n=40$), as depicted in Fig.~\ref{fig:crossval}. Using this cross-validation method ensures that  each validation or test set has been used only once during training in each iteration and never used before in any other iterations. In addition, this method tests the proposed algorithm on all of the 400 images and produces the average performance as a more reliable evaluation metric. 

\begin{figure*}[!t]
\includegraphics[width=1.0\textwidth,trim={1.4cm 0.4cm 0.4cm 0.4cm},clip]{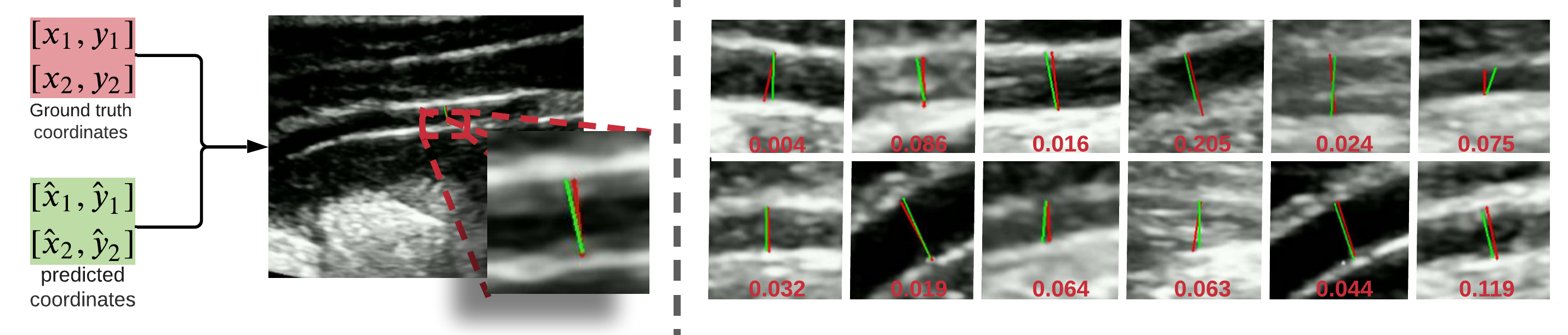}
\centering
\caption{\textbf{Left}: Qualitative results. Here, the prediction results obtained from our novel model with the LCFCN loss and CoordConv layer is shown. \textbf{Right}: Sample prediction results from our model compared with ground-truth labels annotated by human experts. The predictions (green line) \alz{are in reasonably good agreement} with the ground-truth labels (red line). \alz{\acf{MWA} score for each image is shown in red}.}
\label{fig:plot}
\end{figure*}

\subsection{Implementation Details}
Our method uses an Imagenet~\cite{ILSVRC15} pre-trained VGG16 FCN8 network~\cite{long2015fully}. Images were first converted to unsigned 8-bit integers, then resized  to a fixed dimension   $512 \times 512$ pixels and normalized using ImageNet statistics. 
\alz{The predicted measurements are based on these resized images.}
For each image in the training phase, we applied the following: random contrast, random brightness and random gamma transformations. \alz{The training was carried out on an Ubuntu 20.04.1 LTS  machine with an NVIDIA GeForce RTX 2080 TI Graphics Card. The model was trained using PyTorch 1.5 with Python 3.7, a batch size of 1 for 100 epochs.
The inference time per image is 47 milliseconds excluding pre/post-processing steps, which take 12 milliseconds. The overall processing time of  59 milliseconds is well within the requirements of clinical use for automatic muscle measurements.}
ADAM~\cite{kingma2014adam} was used with \alz{ an initial learning rate of $10^{-4}$, and a learning rate decay of $0.1$ after every $30$ epochs.} To overcome the small size problem of the TrA400 dataset, cross-validation was used to report performance across 10 folds, where all images in the dataset are measured using the proposed network and one measurement prediction is generated for each of the 400 images.  

\section{Results}

To fully evaluate our model and compare it with other methods, we provide its experimental results along with a baseline landmark detection approach and a LCFCN approach. Our shown results in Table \ref{tab:Results} are obtained from four fully-supervised networks using point-level annotations. These include

\begin{figure}[!t]
\includegraphics[width=0.48\textwidth]{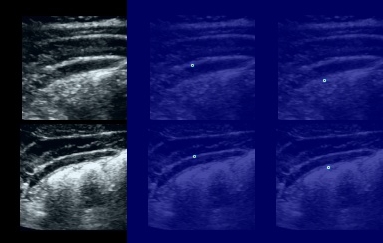}
\centering
\caption{\alz{A sample \ac{US} image (left) and its corresponding predicted scale-aware high-resolution heatmaps \cite{cheng2020higherhrnet} (right).}}
\label{fig:HRNet}
\end{figure}

\begin{itemize}
\setlength\itemsep{0mm}
\item \textbf{Landmark}: 
As a baseline, the standard landmark detection approach has been used similar to the work done in \cite{sofka2017fully}.
With an ImageNet~\cite{ILSVRC15} pre-trained ResNet50~\cite{He2015ResNet} backbone, we regress the point locations using a \ac{FCN} to localize landmarks.

\item \textbf{HigherHRNet}~\cite{cheng2020higherhrnet}: 
\alz{Is a bottom-up method for \ac{2D} human pose estimation that aims to localise human anatomical keypoints. It learns scale-aware representations using a high-resolution feature pyramid. This approach can solve the scale variation challenge in bottom-up multi-person pose estimation and localise key points more precisely. In this study, we have used the same approach to localise the caliper endpoints in \ac{US} images by generating high-resolution spatial heatmaps (see Fig.~\ref{fig:HRNet}).}

\item \textbf{LCFCN} \cite{laradji2018blobs}:
LCFCN is based on a semantic segmentation architecture that is similar to \ac{FCN}~\cite{long2015fully}. LCFCN method optimizes a loss function that ensures that only a single small blob is predicted around the centre of each object (in our case a caliper point) to stop the model from predicting large blobs that merge several object instances. We take the centroid of each blob to get coordinate locations and measure the distances between coordinate pairs to measure the TrA thickness.
\item \textbf{LCFCN+CoordConv}:
We improved LCFCN~\cite{laradji2018blobs} prediction by adding a CoordConv~\cite{liu2018intriguing} layer to the beginning of the FCN8~\cite{long2015fully} architecture. The image is first passed through a CoordConv layer to add pixel-wise spatial location information to it. It then passes through the FCN8 layers to allow the CNN to locate caliper endpoints in the image, more efficiently. 
\end{itemize}

\subsection{Evaluation Procedure}
We evaluate our models using \alz{\acf{MAE} and standard deviation ($\sigma$) across 10 folds,} and \acf{MWA}, as  performance metrics.

\begin{table*}[!t]
\centering
\resizebox{0.8\textwidth}{!}{%
\begin{tabular}{lccc|c|c}
\toprule
           & \multicolumn{3}{c}{MWA} & \multicolumn{1}{c}{\alz{MAE}} &  \multicolumn{1}{c}{\alz{$\sigma$}}\\
\midrule
    Method &  50th & 75th &  95th  &  Average  &     Average  \\
\midrule
    Landmark  & 6.483 & 11.563 &  13.193 & 0.765  &  2.454 \\ 
    \alz{HigherHRNet}  & 4.025 & 4.854 &  7.391 & 0.415  &  1.538 \\ 
    LCFCN   & 3.216 & 4.976 &   6.987  & 0.354  &  1.604\\
    LCFCN + CoordConv    & \textbf{2.710} & \textbf{4.803} &   \textbf{6.641}  & \textbf{0.312}   &  \textbf{1.499}\\\hline

\bottomrule
\end{tabular}
}
\caption{\ac{MWA}, \ac{MAE},  and standard deviation ($\sigma$) metrics for the test sets across all folds. \ac{MWA} was computed by comparing human and \ac{CNN} measurement (\ref{eq:length}) for different percentiles (50th, 75th, 95th). \alz{\ac{MAE} and $\sigma$ are the average across 10 folds.} 
}
\label{tab:Results}
\end{table*}

\alz{
\textbf{\ac{MAE}} measures the deviation of  the predicted and ground-truth endpoints, and is described using (\ref{eq:mae}).
\begin{equation}
\operatorname{MAE}=\frac{1}{N}  \sum_{i=1}^{N}|\hat{E}_i - E_i|,
\label{eq:mae}
\end{equation}
\noindent where $\hat{E}_i$ and and $E_i$ are the predicted and ground truth endpoint coordinates for image $i$, respectively. $N$ is the total number of images considered.
}

\alz{
\textbf{The standard deviation} ($\sigma$)  shows the standard  deviation between predicted and ground-truth endpoints across $10$ folds.}

\textbf{\ac{MWA}}~\cite{sofka2017fully} is a metric used to compare human measurements against \ac{CNN}-predicted measurements. To calculate \ac{MWA}, here we define an error measurement formula that compares the human-annotated and machine-predicted lines connecting two caliper endpoints using (\ref{eq:length})
\begin{equation}
\operatorname{MWA} = \frac{1}{N}\sum_{i=1}^N \abs{ \frac{\left \|\bar{t} - \bar{b}\right \| - \left \|\hat{t} - \hat{b}\right \|}{\left \|\bar{t} -\bar{b}\right \|} } \;,
\label{eq:length}
\end{equation}
where $\bar{t}$ and $\bar{b}$ are top and bottom ground truth caliper endpoints, respectively, and $\hat{t}$ and $\hat{b}$ are top and bottom predicted endpoints, respectively. The error is computed as the sum of the \alz{absolute} length difference between human measurements and CNN predicted measurements. The $50th$, $75th$, and $95th$ error percentiles are tabulated in Table~\ref{tab:Results} to show the distribution of errors and evaluate complex predictions more directly.
\alz{The thickness range of RTrA is between 1.05 and 5.24 mm, while the average is 3.03 mm. The thickness range of CTrA is between 2.07 and 8.02 mm, while the average is 5.25 mm.}
Figure \ref{fig:plot} shows the qualitative results to compare human measurements against \ac{CNN}-predicted measurements.
As the figure demonstrates, the predictions obtained from our proposed network (green line) are in \alz{reasonably} good agreement with the ground-truth labels (red line) annotated by human experts.

Overall, the results summarized in Table \ref{tab:Results} shows that the proposed LCFCN + CoordConv network outperforms previous localization networks, achieving lower  \ac{MAE} and \ac{MWA} on the average of all cross-validation test subsets. It is inspiring that close to expert level performance was achieved with a small training dataset. We believe that the accuracy of our FCNs would increase if they are trained on a larger dataset. Overall, LCFCN with CoordConv layer improves results for all test subsets. The final prediction accuracy at the $50th$ percentile is 2.710\% of the  measurement length which is within the average inter-observer error of 5\% \cite{pirri2019inter,wilson2016measuring}. 

\alz{Statistical significance paired $t$-test was carried out to better compare the human measurements and our proposed method’s predicted values of muscle thickness. Here, our null hypothesis ($H_0$)  was that the true mean difference between the human and our method’s measurements is zero. The alternative hypothesis ($H_A$), on the other hand, was that the true mean difference between the two measurements is not equal to zero. The values used in the two-tailed $P$-value calculations are $t$-score $= 0.1371$, degrees of freedom $= 399$, and mean difference of $0.055$. 
The critical two-tailed $t$-value was $1.9659$ and the calculated $P$-value was $0.891021$. These show that there is not enough evidence to reject the null hypothesis. This decision is made at significance level $\alpha = 0.05$ and $95\%$ confidence interval.
}

\section{Discussion}

The main goal of this study was to investigate various \ac{DL} approaches and find the best method that can facilitate automatic but accurate measurement of the abdominal muscle thickness in \ac{2D} \ac{US} images. An initial naive method that comes to mind is to develop a \ac{DL} network that is trained using image-thickness pairs to predict a thickness value for a given ultrasound image. However, this approach cannot be trusted and explained because there maybe no meaningful relationship between the actual muscle thickness and the network-predicted value and the learned task will not certainly be generalizable to similar abdominal ultrasound images.  
To make sure we have better explainability of the network outcome, our investigations targeted a way that performs the measurement similar to a human operator, i.e. finding key-point locations of the abdominal muscle to predict its thickness. Hence, we opted to treat the task at hand as a landmark detection problem.

Intrigued by the recent success of the \ac{FCN}-based deep learning architecture~\cite{long2015fully} in segmentation problems, we modified this architecture for localization by incorporating LCFCN loss that extend the segmentation loss to perform localization with point supervision~\cite{laradji2018blobs}. Unlike U-net~\cite{ronneberger2015u} that requires the full per-pixel instance segmentation labels, \ac{FCN} uses point-level annotations only. This enabled us to provide localization training points, which in our case are caliper points on an abdominal muscle. 
We also added a CoordConv~\cite{liu2018intriguing} layer, which introduces two extra channels to the network and enables the CNN to learn the order of convolutional filters through coordinates, in order to properly learn translation equivairance~\cite{liu2018intriguing}.

After extensively tuning our customized network for the measurement task, our localization \ac{FCN} achieved competitive localization performance in the experiments even with a limited training dataset.
We evaluated the localization performance of our proposed \ac{FCN} method on 10 folds test set that contains \ac{2D} \ac{US} abdominal muscle images of different anatomical structures, demonstrating the generic applicability of our proposed method. 
Our results show that when training on ($n=320$) training set images, our method achieved close to expert ultrasound technicians level performance \cite{pirri2019inter,wilson2016measuring}.
We only used a limited number of training set images. A greater number of training set images could lead to a higher generalization of our results.

The 10-fold cross-validation technique used in this paper is one way to overcome the limited number of dataset images and improve prediction credibility over the holdout method. Firstly, the dataset is divided into 10 subsets, and the holdout method is repeated 10 times (see Fig.~\ref{fig:crossval}). 
Secondly, each time, one of the 10 subsets is used as the test set and the other 9 subsets are used to form training and validation sets. 
Finally, the average error across all 10 predictions is computed. 
The prediction for each single test subsets was consistent with the average of all cross-validation test subsets. This confirms that our model appropriately generalizes to new data.

A limitation of this study are that all the still images were extracted only from eligible participants, in a single site and with a single acquisition protocol~\cite{kennedy2019intra}. To avoid overfitting to a single acquisition protocol, exams should be obtained on several sites.
Thus, future studies are required to ensure the following: 
\begin{itemize}
\item The dataset to be collected from several sites; 
\item All measurements should be performed by a single technician experienced in \ac{2D} \ac{US} measurements;
\item The \ac{US} images to be labelled multiple times by the same expert;
\item Consistent \ac{US} image quality is achieved in all measurements.
\end{itemize}

As mentioned in Section \ref{sec:relwork}, so far no one appears to have applied \ac{FCN} variants such as an \alz{LCFCN or LCFCN+CoordConv} for US image muscle thickness measurements.
The importance of our results on using such networks thus lies both in their generality and their relative ease of application to automatic measurements of ultrasound images. We hope that our contribution can be used and extended for other measurement tasks across other medical imaging domains where a localization and/or measurement task is at hand.

\section{Conclusion}
In this work, we used the \ac{DL} localization approach to automate measuring abdominal muscle dimensions in \ac{2D} ultrasound imaging.
We proposed to detect measurement key-point locations by computing their localized estimates with a \ac{FCN}. To improve localization performance, we added a CoordConv layer to \ac{FCN} and incorporated the \ac{LCFCN} loss function. This novel design resulted in errors below 3\% of \ac{MWA} between the measurements by human and the \ac{CNN} predictions, which is within normal inter-observer variability.  
As this measurement is one of the outcome measures used to evaluate abdominal muscle dimensions to aid rehabilitation, our developed tool can assist by automatic measurements of ultrasound images, while reducing the inter-rater variability. This tool can provide more reliable diagnostic and
outcome measurements, benefiting both patients and healthcare professional.

\section*{Acknowledgement}
This research is supported by an Australian Research Training Program (RTP) Scholarship. 
MRA would like to thank Dr Carla Ewels for fruitful discussion regarding the statistical significance of the proposed method compared to human gold standard.

\bibliographystyle{IEEEtran}
\bibliography{References}

\end{document}